# Phase Change Control of Interlayer Exchange Coupling

Xiaofei Fan,[1,†] Guodong Wei,[1,†] Xiaoyang Lin,[1,2,†,*] Xinhe Wang,[1] Zhizhong Si,[1] Xueying Zhang,[1,2] Qiming Shao,[3] Stephane Mangin,[4] Eric Fullerton,[5] Lei Jiang,[6] Weisheng Zhao[1,2,*]

[1] Fert Beijing Research Institute, School of Microelectronics & Beijing Advanced Innovation Center for Big Data and Brain Computing (BDBC), Beihang University, Beijing 100191, China

[2] Beihang-Goertek Joint Microelectronics Institute, Qingdao Research Institute, Beihang University, Qingdao 266000, China

[3] Department of Electronic and Computer Engineering, The Hong Kong University of Science and Technology, Clear Water Bay, Kowloon, Hong Kong SAR, China

[4] Institut Jean Lamour, UMR 7198, CNRS-Universite de Lorraine, F-54000 Nancy, France

[5] Center for Memory and Recording Research, University of California San Diego, 9500 Gilman Drive, La Jolla, CA 92093-0401, USA

[6] School of Chemistry, Beihang University, Beijing 100191, China

†These authors contributed equally.

*E-mail: XYLin@buaa.edu.cn (X.Y.L), weisheng.zhao@buaa.edu.cn (W.S.Z)

**Abstract**

Changing the interlayer exchange coupling between magnetic layers *in-situ* is a key issue of spintronics, as it allows for the optimization of properties that are desirable for applications, including magnetic sensing and memory. In this paper, we utilize the phase change material $VO_2$ as a spacer layer to regulate the interlayer exchange coupling between ferromagnetic layers with perpendicular magnetic anisotropy. The successful growth of ultra-thin (several nanometres) $VO_2$ films is realized by sputtering at room temperature, which further enables the fabrication of $[Pt/Co]_2/VO_2/[Co/Pt]_2$ multilayers with distinct interfaces. Such a magnetic multilayer exhibits an evolution from antiferromagnetic coupling to ferromagnetic coupling as the $VO_2$ undergoes a phase change. The underlying mechanism originates from the change in the electronic structure of the spacer layer from an insulating to a metallic state. As a demonstration of phase change spintronics, this work may reveal the great potential of material innovations for next-generation spintronics.

**Introduction**

Interlayer exchange coupling (IEC) is an indirect exchange interaction between magnetic layers mediated by the spacer layer. The discovery of IEC has given birth to a boom in spintronics research, including the study of giant magnetoresistance[1,2] and tunnelling magnetoresistance effects[3,4], and the design of synthetic antiferromagnetic layers in magnetic tunnel junctions[5]. Different from direct manipulations of the magnetization in spintronic devices by magnetic field and spin torques[6–8], the IEC switch between ferromagnetic (FM) and antiferromagnetic (AFM) coupling states may provide a promising opportunity to realize low-power data write-in in the post-Moore's-law era[9–14].

To date, IEC has been systematically investigated in magnetic multilayers with different metallic spacer layers[15–17] as well as some semiconducting and insulating spacers[18–22]. For IEC through metallic spacer layer, the periodical oscillation of coupling strength with the spacer layer thickness[15,16,23] has been explained by Rudermann–Kittel–Kasuya–Yosida (RKKY) theory[24–26], which originates from the interactions between localized d- or f-orbit electrons via the conduction electrons. Meanwhile, band excitations have been used to explain the IEC mediated by semiconductors[27,28]. For insulating spacers, IEC with features different from those of metallic spacer layers has been interpreted by the spin dependent tunnelling[29,30]. Pioneering theoretical studies have also been performed to develop a unified theory for kinds of spacer layers by introducing the concept of a complex Fermi surface[31,32]. Experimental explorations of IEC through spacer layers with tuneable electrical properties are thus of both fundamental and practical significances[14,33]. From this point of view, phase change materials with metal-to-insulator transition (MIT) features[34,35], could be a promising candidate of emerging spacer layer for IEC regulations.

In this work, we study the IEC evolution in magnetic multilayers with a phase change spacer layer $VO_2$ which features of near room temperature MIT and fast-response speed[36,37]. The deposition of ultra-thin (several nanometres) $VO_2$ films is realized by magnetron sputtering in high vacuum. The MIT feature of this ultra-thin film is further proved by electrical transport measurements. Magnetic multilayer samples of [Pt/Co]$_2$/VO$_2$/[Co/Pt]$_2$ with perpendicular magnetic anisotropy (PMA) are prepared to explore the phase-change-induced regulation of the IEC. Switching from AFM coupling with insulating VO$_2$ to FM coupling with metallic VO$_2$ is observed. The IEC regulation behaviour is further explained by the effect of MIT-induced Fermi surface modulation, changing the coupling mechanism from impurity-assisted spin dependent tunnelling to RKKY-type interaction.

**Results**

**Characterization of ultra-thin VO$_2$ film and preparation of phase change magnetic multilayers**

For the experiment of the phase-change-induced IEC regulation, the preparation of VO$_2$ films with nanometre thicknesses is fundamental to the fabrication of VO$_2$-based magnetic multilayers. Both spin dependent tunnelling and RKKY induced IEC result in an exponential decay of the coupling strength, which usually confines the critical thickness of the spacer layer below several nanometres[15,38]. Although many methods have been used to deposit VO$_2$, such as molecular beam epitaxy[39], pulsed laser deposition[40] and chemical vapour deposition[41], strict growth conditions, *e.g.*, specific substrates, high temperature and high oxygen pressure, limit their applications in spintronics[42,43]. A deposition method with abilities of nanometre-accuracy, multilayer integration and interface engineering is thus desired.

We choose magnetron sputtering as the deposition method, due to its ability to achieve multilayer deposition in ultra-high vacuum (UHV) with sub-nanometre thickness accuracy. Through a careful optimization of the target component and a systematic study of the deposition conditions, ultra-thin VO$_2$ films with thicknesses ranging from several angstroms to several nanometres can be successfully prepared on different substrates in UHV at room temperature. A broad peak, indicating their amorphous feature, has been found in the X-ray diffraction (XRD) result (see Supplementary **Fig. S1**). X-ray photoelectron spectroscopy (XPS) measurement was applied to check the valence states of vanadium. As shown in **Fig. 1a**, the proportion of VO$_2$ reaches more than 89.3%, which demonstrates the relatively pure composition of the film. To directly exam the phase change feature of the VO$_2$ film, the resistance of a vertical tunnel junction device fabricated based on Au/VO$_2$ (2 nm)/Au multilayers is measured at different temperatures. The I-V curves with increasing temperature are shown in **Fig. 1b**, from which an evolution from a tunnelling contact to a transparent contact can be detected. The results indicate that the resistance of the VO$_2$ spacer layer exhibits a significant decrease after heating; *i.e.*, an MIT occurs. **Fig. 1c** presents the ratio of the resistance change (calculated as the resistance divided by its minimum value) versus temperature with a measured current of 10 nA for the 2-nm-thick sample and a 40-nm-thick single-crystal film sample. It is noteworthy that the phase change in the nanometre-thick VO$_2$ layer occurs at a temperature of approximately 310 K, which is much lower than that of the thicker sample. This difference is probably caused by the effect of interfacial strain, which has also been observed in single-crystal samples[44]. According to some reports, the MIT feature of ultra-thin VO$_2$ samples is usually suppressed due to impurity diffusion from the substrate[45], which could be largely restrained by our room-temperature UHV deposition condition. Although potential factors such as the amorphous nature and inevitable defects still affect the MIT amplitude (~20 times of the resistance change), the hysteresis phenomenon proves the existence of the phase change in the ultra-thin film.

The appreciable MIT effect, good compatibility of the substrates, and UHV room-temperature deposition conditions of the ultra-thin VO$_2$ film guarantee the possibility of fabricating VO$_2$-based magnetic multilayers. To investigate the effect of the phase change regulation of the IEC, we prepared a series of [Pt/Co]$_2$/VO$_2$/[Co/Pt]$_2$ heterostructures on SiO$_2$/Si substrates (see **Fig. 1d** for a schematic diagram of the multilayers). VO$_2$ layers with different thicknesses are used as the spacer layers, and Co/Pt systems are chosen as the ferromagnetic layers with PMA in these samples. The interface quality is checked by high-resolution transmission electron microscopy (HRTEM), by which the thicknesses of the spacer layers are measured to be 0.76, 1.48, 1.83 and 2.26 nm (see Supplementary **Fig. S2**). Taking the multilayer with the 2.26-nm-thick VO$_2$ spacer layer as an example (**Fig. 1e**), distinct interfaces and good continuity between the spacer and the FM layers can be observed. However, some inhomogeneous region with microcrystalline morphology can also be observed within the VO$_2$ layer, which may be caused by the fluctuations of the sputtering energy and non-annealing process. These inevitable defects can explain the sluggish MIT curve and non-uniform phase change process mentioned in the following discussion.

**Fig. 1f** presents a schematic diagram of the modulation principle. In the VO$_2$-based magnetic multilayer, the magnetic electrons in the top and bottom layers are coupled indirectly through VO$_2$, whose electron density of 3d orbitals vary tremendously in the phase change process. As reported in different oxide space layers[18,29], spin dependent tunnelling effect can dominate the coupling through insulating VO$_2$ at room temperature. When VO$_2$ is conducting at high temperature, the RKKY-type coupling mediated by conduction electrons plays the dominant role. As a result, the coupling strength or even coupling type can be changed owing to the electron density enhancement near the Fermi surface. In the following study, we find that the coupling is AFM at room temperature and evolves into FM coupling after the VO$_2$ MIT occurs. A detailed analysis of the change in the electronic structure will be discussed later in this article.

**Phase-change-induced IEC regulation**

The IEC regulation in the phase change magnetic multilayer is first investigated by magnetic property measurements. A vibrating sample magnetometer (VSM) and the polar magneto-optic Kerr effect (p-MOKE) are used to characterize the [Pt/Co]$_2$/VO$_2$/[Co/Pt]$_2$ samples. As shown in **Fig. 2a**, obvious anisotropy is detected in the sample with a 0.76-nm-thick VO$_2$ spacer layer. Similar PMA properties can be found in all the other films (see Supplementary **Fig. S3**). The magnetization shows two reorientation processes when an out-of-plane magnetic field is applied, which indicates strong FM coupling within both the top and bottom Co/Pt bilayers. Thus, these bilayers can be assumed to be two single FM units when the IEC regulation is investigated.

As illustrated in **Fig. 2b-2e** hysteresis loops are also measured by the p-MOKE at room temperature (RT, 300 K) and high temperature (HT, 360 K) for samples with different thicknesses of VO$_2$. Interestingly, the two magnetization flips merge into one after the MIT occurs in the results of the 0.76-, 1.48-, 1.83- and 2.26-nm-thick VO$_2$ samples. This simultaneous switching phenomenon, implying strong FM coupling between the top and bottom Co/Pt bilayers through the metallic $VO_2$, provides solid evidence that the IEC between magnetic layers has been changed. The temperature dependent experimental results of the control samples (with the structure of [Pt/Co]$_2$/$VO_2$ and $VO_2$/[Co/Pt]$_2$, Supplementary **Fig. S4**) show that the magnetization reorientation of two control samples always happen at different magnetic fields. Additionally, the in-plane hysteresis loops of 0.76-nm-thick $VO_2$ sample coincide with each other at different temperatures, implying no obvious anisotropy change during the phase change process. (**Fig. 2f** and Supplementary **Fig. S7**). These features, which exclude the possibility of differentiated PMA modulation effect on the top and bottom magnetic layer, confirm the existence of IEC regulation via the MIT of $VO_2$.

To further understand the phase-change-induced IEC regulation, the first priority is to distinguish the magnetic coupling state at RT for the samples in which the regulation can be observed. Taking the sample with 0.76-nm-thick $VO_2$ as an example, a high-accuracy VSM with a built-in Hall probe (accuracy of 0.01 Oe) is applied to collect the minor loop information of the top magnetic layer. As shown in the inset of Fig. 3a, a positive exchange bias field $H_{ex-top}$ (~0.35 Oe, see **Fig. 3a**) can be detected in multiple minor loop measurements, which indicates AFM coupling at RT. To further confirm this result, the magnetic domain switching behaviour around the field region of the minor loop is also investigated, as shown in **Fig. 3b**. The results verify that the domain switching fields of the top magnetic layer decrease near -0.2 Oe and 1.88 Oe. Meanwhile, similar minor loop measurements are carried out for samples with various thicknesses of $VO_2$ (see Supplementary **Fig. S5**), and the results suggest that the samples with $VO_2$ thicknesses of 1.48, 1.83 and 2.26 nm also exhibit AFM coupling ($H_{ex-top}$=0.64, 0.30, and 0.15 Oe, respectively) at RT, while the sample with the thickness of $VO_2$ greater than 3nm exhibits a decoupling feature.

To obtain more details of the regulation effect, we focus on the dynamic change of the IEC effect. **Fig. 3c** shows the hysteresis loops of the samples at different temperatures. In all the four samples, there are double flips in the loops with the insulating $VO_2$. As the temperature increasing, we find that the loop features a smooth flip of the top FM layer before a sudden transition into one single flip (FM coupling with metallic $VO_2$). Interestingly, the critical temperature of the change in the IEC type shows obvious differences. For the thinner two samples, the FM coupling change occurs above RT (approximately 320~330 K), while for the thicker two samples, the FM coupling change occurs at approximately 310 K. Since all the samples are deposited at RT with the same growth parameters, one possible reason is the appreciable tuning of MIT temperature by interfacial strength at nano-scale, which has also been observed in single-crystal ultrathin $VO_2$ samples[36].

Considering that the phase change of $VO_2$ exhibits obvious hysteresis (**Fig. 1d**), it is reasonable to believe that a phase-change-induced IEC regulation should exhibit the same feature. Therefore, we measured the hysteresis loops of the 0.76-nm-thick $VO_2$ sample at 305 and 315 K during the heating-up and cooling-down process. As shown in **Fig. 3d**, the sharpness of the flip in both results shows a clear difference, which proves our speculation regarding the hysteresis and further verifies that the coupling change is indeed caused by a phase change in $VO_2$.

Based on the results obtained from the $[Pt/Co]_2/VO_2/[Co/Pt]_2$ multilayer samples, the exchange bias field between two FM layers can be derived from the center shift of minor loops (for example the $H_{ex\text{-}top}$ in Fig. 3a ) or the shift of the flip edges relative to those of a control sample[21]. Considering the canted shape of the top layer minor loop at HT, we use the flip edge of the bottom layer ($H_{ex\text{-}bottom}$) to calculate the value of exchange bias field, so that the standard can be kept consistent in the whole temperature range. As plotted in **Fig. 4a**, the negative $H_{ex\text{-}bottom}$ values represent AFM coupling, and the positive values represent FM coupling. Samples with various thicknesses of $VO_2$ all demonstrate a transition from AFM coupling at a lower temperature (with insulating $VO_2$) to stronger FM coupling at a higher temperature (with metallic $VO_2$). For each of the four samples, a critical temperature exists, corresponding the change from AFM to FM coupling.

**Discussion**

The $H_{ex\text{-}bottom}$ value, which reflects the coupling strength, is found to decay exponentially at RT with thicknesses increasing of the spacer layer. (see Supplementary **Fig. S6**) This behaviour is consistent with the spin dependent tunnelling induced coupling as suggested for insulating spacer layers[18,21,29]. Notably, the data also shows a comparable coupling strength at AFM state (-1~0 Oe) to that of the FM state (0~2.5 Oe). This relatively strong coupling can be explained by the impurity-assisted mode[46]. As the phase change occurs, the increased conduction electron density contributes to IEC and the coupling change into the RKKY mode. However, the disappearance of oscillating IEC in the VO$_2$ metallic state is probably due to the restricted range of the VO$_2$ thickness or an inherent feature of such phase change spacer mediated IEC.

To explain the mechanism behind the transition, the change in the electron state density and Fermi surface of the spacer layer should be taken into consideration. As elucidated in previous research[47,48], the Peierl-like phase change can produce a large spectral weight transfer from π* ($d_{xy}$) orbitals into the valence $d_{//}$ ($d_{xz,\,yz}$) orbitals (**Fig. 4b**). For the MIT of VO$_2$ induced by a temperature change in our experiments, this enhancement of d-orbital electrons will then influence the coupling between ferromagnetic layers. In fact, calculation results have been reported for Co-doped TiO$_2$/VO$_2$ diluted magnetic semiconductor multilayers, showing that an FM-AFM coupling change may accompany the phase change in VO$_2$[49]. As our film structure and spin orientation are much more complicated than an ideal model, further calculations and simulations are needed to reveal more details of the physics.

Besides the AFM to FM transition, the variation of the loop shape in the phase change process also provides a clue to understand the dynamic spin evolution process. If the phase change in VO$_2$ is uniform across the whole spacer layer, the IEC regulation will make the flips of magnetic layers approach each other at the FM coupling region, as illustrated in the uniform IEC change route in **Fig. 4b**. However, for the evolution process in our results, the coupling change is accompanied by the smoothing behaviour for the first magnetization flip. As a macroscopic presentation of non-uniform switching, this behaviour may be related to the inhomogeneous feature of the amorphous ultra-thin VO$_2$ film[50,51], which could induce domains with different conductivity during the phase change. Thus, the coupling also becomes non-uniform in the conducting and insulting areas.

To confirm this speculation, a Kerr microscope is used to observe the domain switching at different temperatures. As illustrated in **Fig. 4c**, a magnetic field is applied to the multilayers with 0.76-nm-thick $VO_2$ to induce a critical flipping state at RT. As the temperature increases, the magnetic domain first evolves into a maze morphology at 310 K and then finely breaks into smaller domains that are beyond the resolution of the microscope at 315 K. With the in-plane loop measurement at different temperatures excluding the possibility of an anisotropy change (Fig. 2f and Supplementary **Fig. S7**), it is reasonable to believe that this change in the domain type is caused by the non-uniform phase change in $VO_2$, which provides more nucleation points for spin switching.

In conclusion, we have succeeded in preparing an ultra-thin $VO_2$ film with an appreciable metal-to-insulator transition by ultra-high-vacuum magnetron sputtering at room temperature. The as-prepared nanometre-thick $VO_2$ has been adopted as a spacer layer to regulate the interlayer exchange coupling in a $[Pt/Co]_2/VO_2/[Co/Pt]_2$ system, whose coupling strength and type can be modulated reversibly and repeatably through the phase change in $VO_2$. Further analyses indicate that changes in the electron state density and shift of the Fermi surface may explain the IEC change. Non-uniform phase-change-induced local magnetic domain switching is used to explain the dynamic spin evolution during the transition process. The current result, which combines spintronics and phase change electronics, offers a new strategy to regulate IEC and provides great possibilities for developing new types of electronic devices.

**Methods**

**Film deposition and device fabrication**

The Au/VO$_2$/Au and [Pt/Co]$_2$/VO$_2$/[Co/Pt]$_2$ multilayers were both grown on SiO$_2$/Si substrates by magnetron sputtering at RT in high vacuum. A stoichiometric target was used to deposit the VO$_2$ film during the experiments. The tunnel junctions were patterned by optical lithography (Micro Writer ML Baby, Durham Magneto Optics) followed by argon ion-beam etching of the bottom electrodes and nanopillars. Then, the samples were fully covered with SiO$_2$ for insulation. After the lift-off procedure, holes were produced over the bottom electrodes. Both the bottom electrodes and tunnel junctions were then connected to 90-nm Ti/Au electrodes using e-beam evaporation to allow electrical contact for the measurements.

**Characterization and measurement**

The magnetic properties were measured by the MOKE (NanoMOKE3, Durham Magneto optics ltd) with a homemade heat source and a VSM (7400, Lakeshore) at different temperatures. The Au/VO$_2$/Au tunnel junctions were measured by normal 4-terminal methods with Keithley 6221 and Keithley 2182 source and measurements units, respectively. HRTEM was performed by a Talos F200X instrument. The XPS results were obtained by a Rigaku D/max-2500PC instrument, and the XRD measurements were performed by a Delta-X instrument.

**Figures & Legends:**

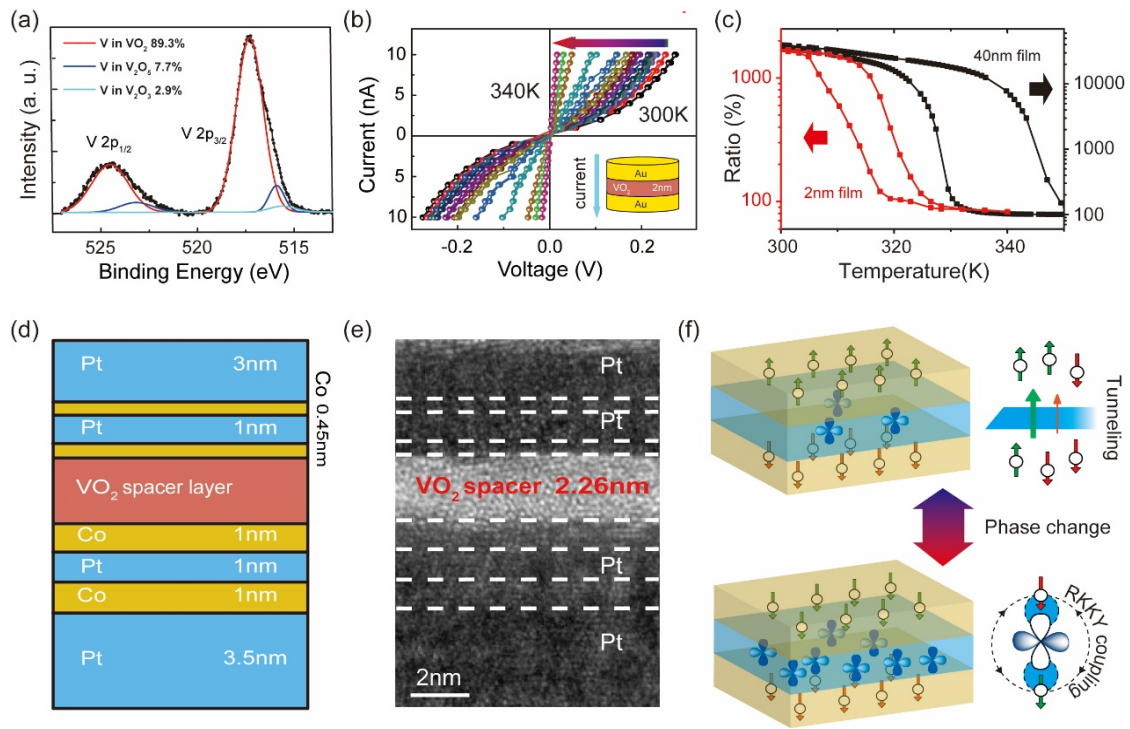

**Figure 1 | Characterization of 2-nm-thick VO$_2$ film grown by sputtering.** (**a**) X-ray photoelectron spectroscopy (XPS) profile of the sample with peak fitting results. The proportion of VO$_2$ reaches 89.3%. (**b**) I-V curve of the VO$_2$ tunnel junction measured at different temperatures with the current flows vertically through the device. The insert gives the schematic of the device which is fabricated based on Au/VO$_2$ (2 nm)/Au multilayer with a diameter of 10 μm. (**c**) The resistance change ratio versus temperature of 2 nm and 40 nm VO$_2$ films with a current of 10 nA, which indicates the existence of MIT feature. The resistance change ratio is defined as the resistance divided by its minimum value. (**d**) The schematic of the [Pt/Co]$_2$/VO$_2$/[Co/Pt]$_2$ multilayer. (**e**) High-resolution transmission electron microscopy (HRTEM) image of the 2.26-nm-thick VO$_2$ spacer sample that shows distinct interfaces with good continuity between the spacer and the ferromagnetic layers. (**f**) Schematic diagram of the interlayer exchange coupling (IEC) regulation principle. The phase change from insulating to metallic can enhance the electron density of 3d orbitals near Fermi surface, making the coupling change from antiferromagnetic (AFM) to ferromagnetic (FM). The corresponding coupling mechanism also changes from the impurity-assisted spin dependent tunnelling (the diagram above) to the RKKY-type interaction (the diagram below).

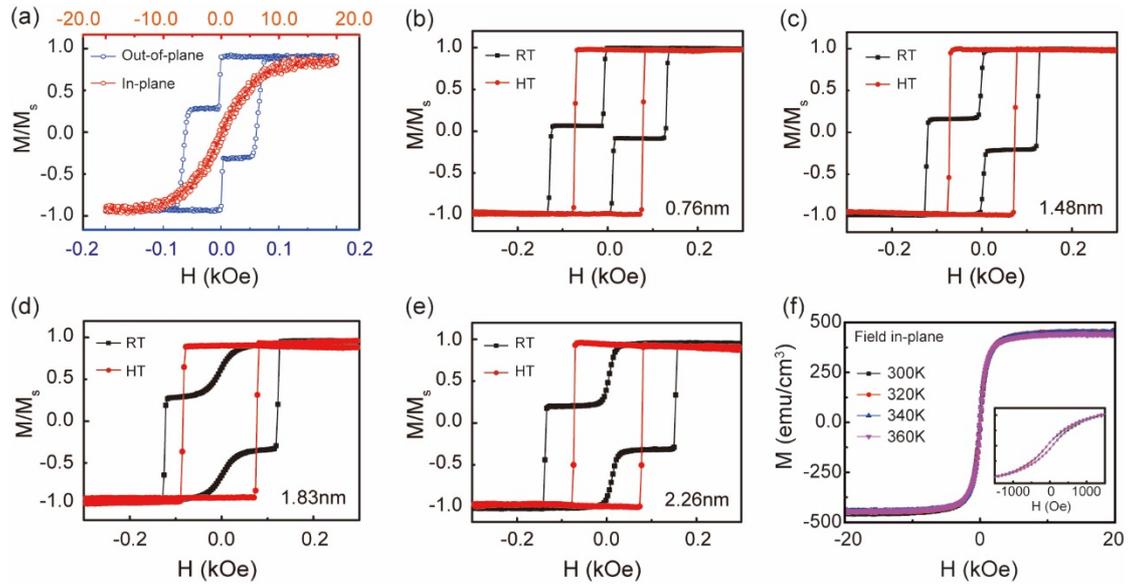

**Figure 2 | Magnetic properties of [Pt/Co]$_2$/VO$_2$/[Co/Pt]$_2$ multilayers.** (**a**) Hysteresis loops of [Pt/Co]$_2$/VO$_2$ (0.76 nm)/[Co/Pt]$_2$ multilayer measured by vibrating sample magnetometer (VSM) with magnetic field applied in-plane field (red line) and out-of-plane field (blue line) at room temperature which shows the PMA feature of the multilayer. *M*, magnetization of the magnetic layers; $M_s$, saturation magnetization of the magnetic layers; *H*, external magnetic field. (**b**)-(**e**) Hysteresis loops of the multilayers measured by polar magneto-optic Kerr effect (p-MOKE) at room temperature (RT, 300 K, black line) and high temperature (HT, 360 K, red line), where the thickness of VO$_2$ is 0.76, 1.48, 1.83 and 2.26 nm, respectively. The two magnetization flips merge into one after the MIT occurs. (**f**) In-plane hysteresis loops of 0.76-nm-thick VO$_2$ sample at different temperatures, which show no indication of anisotropy change in the phase change process. The inset gives the enlarged image from -1000 to 1000 Oe.

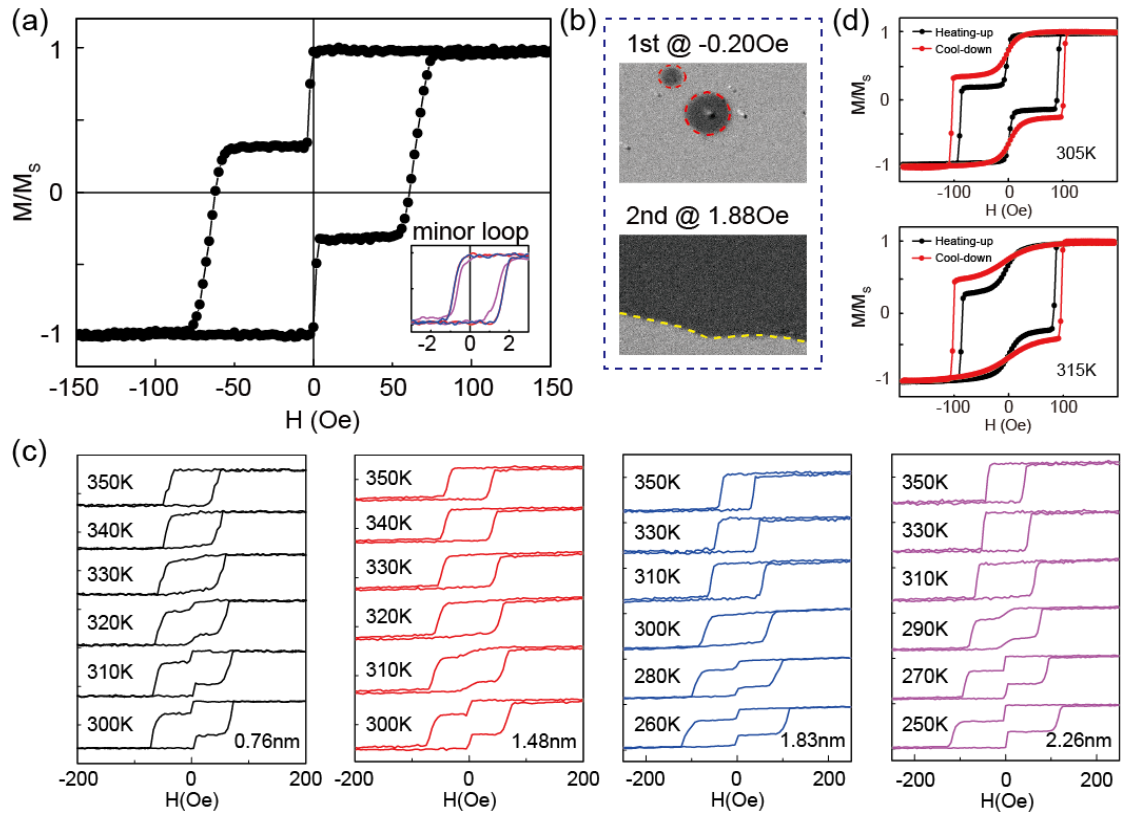

**Figure 3 | Regulation of IEC in the multilayers during phase change process.** (**a**) Major loop and minor loops of the multilayer with 0.76-nm-thick $VO_2$ measured by VSM. The inset gives the minor loops obtained in the magnetic range of ±3 Oe with stepping of 0.1 Oe. (**b**) Magnetic domain switching behaviour of the sample around the minor loop field range obtained from Kerr microscope with out-of-plane field applied. The centre deviation of minor loop flips from zero field to positive field observed in (a) and (b) indicates an AFM coupling exists. (**c**) The variation of hysteresis loops measured by VSM as temperature rising for samples with $VO_2$ of 0.76, 1.48, 1.83, 2.26 nm respectively. The loop features a smooth flip of the top FM layer before a sudden transition into one single flip, and the critical temperatures of IEC type change show obvious difference. (**d**) The comparison of the hysteresis loops between heating-up and cool-down process at 305 K and 315 K. The sharpness of the flip in both results shows a clear difference which is consistent with $VO_2$ hysteresis feature.

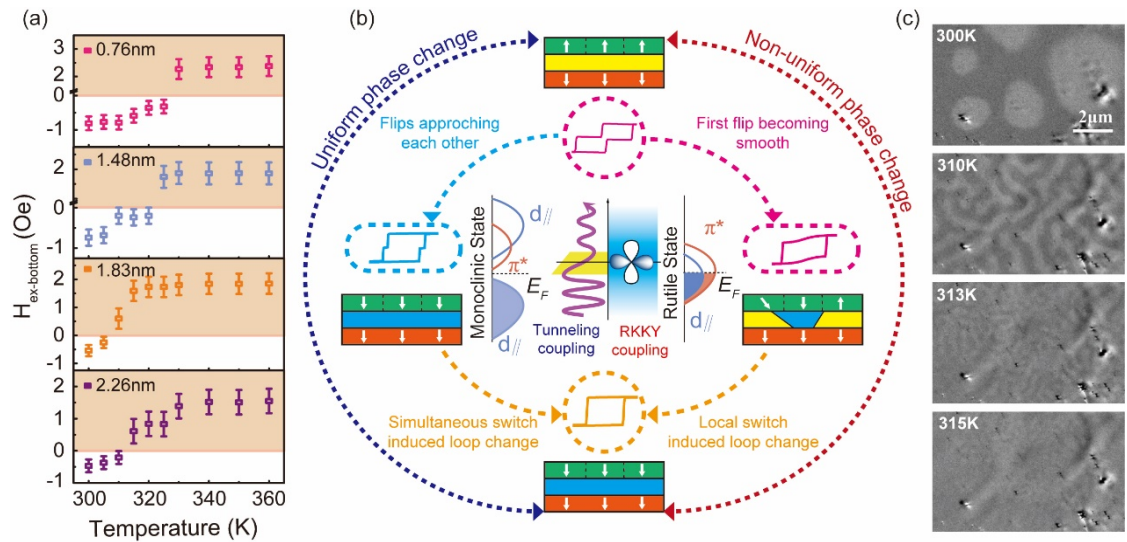

**Figure 4 | Physical mechanism of IEC regulation via phase change.** (**a**) Exchange coupling field of bottom magnetic layer ($H_{ex\text{-}bottom}$) versus temperature for $VO_2$ thicknesses of 0.76, 1.48, 1.83 and 2.26 nm, respectively. (**b**) Schematic diagram of the IEC regulation mechanism. The inner image gives the energy band variation of $VO_2$ and corresponding coupling mechanism before and after phase change. The spectral weight transfer from $\pi^*$ ($d_{xy}$) orbitals into the valence $d_{//}$ ($d_{xz, yz}$) orbitals enhances the electron density near Fermi surface, and makes coupling mechanism change from the spin dependent tunnelling to the RKKY-type interaction. The outside gives the different dynamic spin evolution process with uniform (left routine) and non-uniform (right routine) phase change. In the non-uniform routine, the electric domain in $VO_2$ may induce local magnetic domain switching during the phase change. (**c**) The magnetic domain morphology of 0.76-nm-thick $VO_2$ film observed by Kerr microscope at different temperature with the field of 2.6 Oe applied. The magnetic domain evolves from bubble type into maze morphology and then breaks into smaller domains that are beyond the resolution of the microscope.